\documentclass[
    aps,
    prl,
    twocolumn,
    10pt,
    longbibliography,
    floatfix,
    superscriptaddress,
    preprintnumbers
]{revtex4-2}

\pdfoutput=1

\usepackage{amsmath}
\usepackage{amsfonts}
\usepackage{amssymb}
\usepackage{amsthm}
\usepackage{amstext}
\usepackage{array}
\usepackage{booktabs}
\usepackage{bm}
\usepackage{calc}
\usepackage{multirow}
\usepackage{textcomp}
\usepackage{txfonts}
\usepackage{ulem}
\usepackage{url}
\usepackage{graphicx}
\usepackage[svgnames,psnames]{xcolor}
\usepackage{hyperref}
\hypersetup{
    colorlinks=true,
    citecolor=DarkGreen,
    linkcolor=FireBrick,
    linktocpage,
    unicode,
}

\definecolor{CiteColor}{rgb}{0,0.5,0}
\hypersetup{citecolor=CiteColor}

\newcommand{\rmd}{{\rm d}}
\newcommand{\lmkn}{{\ell m k n}}

\definecolor{red  }{rgb}{1,0,0}
\definecolor{blue }{rgb}{0,0,1}
\definecolor{green}{rgb}{0,1,0}

\begin{document}

\title{Adiabatic waveforms from extreme-mass-ratio inspirals: an analytical approach}

\def\addSoton{School of Mathematics and STAG Research Centre, University of Southampton, Southampton SO17 1BJ, United Kingdom}
\def\addIIP{International Institute of Physics, Universidade Federal do Rio Grande do Norte, Campus Universit{\'a}rio, Lagoa Nova, Natal-RN 59078-970, Brazil}
\def\addCaltec{Theoretical  Astrophysics  Group,  California  Institute  of  Technology,  Pasadena,  CA  91125,  USA}
\def\addRyukoku{Faculty of Law, Ryukoku University, Kyoto 612-8577, Japan}
\def\addYITPGW{Center for Gravitational Physics, Yukawa Institute 
for Theoretical Physics, Kyoto University, Kyoto 606-8502, Japan}
\def\addKyoto{Department of Physics, Kyoto University, Kyoto 606-8502, Japan}
\def\addOsaka{Advanced Mathematical Institute, Osaka City University, Osaka 558-8585, Japan}
\def\addOtemon{Institute of Liberal Arts, Otemon Gakuin University, Osaka 567-8502, Japan}

\author{Soichiro~Isoyama}
\affiliation{\addSoton}
\affiliation{\addIIP}

\author{Ryuichi~Fujita}
\affiliation{\addOtemon}
\affiliation{\addYITPGW}

\author{Alvin~J.~K.~Chua}
\affiliation{\addCaltec}

\author{Hiroyuki~Nakano}
\affiliation{\addRyukoku}

\author{Adam~Pound}
\affiliation{\addSoton}

\author{Norichika~Sago}
\affiliation{\addKyoto}
\affiliation{\addOsaka}

\date{\today}

\preprint{YITP-21-114, KUNS-2901, OCU-PHYS-551, AP-GR-175}

\begin{abstract}
Scientific analysis for the gravitational-wave detector LISA will require theoretical waveforms from extreme-mass-ratio inspirals (EMRIs) that extensively cover all possible orbital and spin configurations around astrophysical Kerr black holes. However, on-the-fly calculations of these waveforms have not yet overcome the high dimensionality of the parameter space. To confront this challenge, we present a user-ready EMRI waveform model for generic (eccentric and inclined) orbits in Kerr spacetime, using an analytical self-force approach. Our model accurately covers all EMRIs with arbitrary inclination and black hole spin, up to modest eccentricity ($\lesssim 0.3$) and separation ($\gtrsim2$--$10M$ from the last stable orbit). In that regime, our waveforms are accurate at the leading `adiabatic' order, and they approximately capture transient self-force resonances that significantly impact the gravitational-wave phase. The model fills an urgent need for extensive waveforms in ongoing data-analysis studies, and its individual components will continue to be useful in future science-adequate waveforms.  
\end{abstract}

\maketitle

{\it Introduction}.---Gravitational-wave (GW) astronomy has revealed a cosmos brimming with black holes (BHs)~\cite{LIGOScientific:2018mvr,LIGOScientific:2020ibl,LIGOScientific:2021usb,LIGOScientific:2021djp}, and 
as GW detectors improve, we will continue to learn more about BHs' properties and demographics. In the 2030s, GW detectors in space (LISA, the Laser Interferometer Space Antenna~\cite{Audley:2017drz}, as well as DECIGO~\cite{Kawamura:2020pcg}, 
TianQin, and Taiji~\cite{TianQin:2020hid,2021PTEP.2021eA108L,Gong:2021gvw}) 
will observe unparalleled probes of BHs: the inspirals of stellar-mass objects into supermassive BHs in galactic cores~\cite{Amaro-Seoane:2012lgq,Amaro-Seoane:2020zbo}. 
The waveforms from these extreme-mass-ratio inspirals (EMRIs, with mass ratios $\eta\sim 10^{-4}$--$10^{-7}$) will contain a unique wealth of information about the spacetime geometry of BHs, strong-field physics in their vicinity, and the astrophysics of their stellar environments~\cite{Amaro-Seoane:2007osp,Babak:2017tow,Berry:2019wgg,Fan:2020zhy,Zi:2021pdp}.

The scientific potential of EMRIs has motivated the community 
to solve the relativistic two-body problem in the small-mass-ratio regime, 
making use of gravitational self-force (GSF)  theory~\cite{Mino:1996nk,Mino:1997bw,Quinn:1996am,Gralla:2008fg,Pound:2009sm,Pound:2012nt,Gralla:2012db}:
a perturbative method in which the small body perturbs the central BH's spacetime, and the perturbation drives the body away from geodesic motion. 
GSF theory has flourished over the past 25 years~\cite{Shah:2012gu,vandeMeent:2016pee,vandeMeent:2017bcc,Pound:2019lzj,Warburton:2021kwk}, 
yielding a range of powerful tools for modelling EMRIs~\cite{Mino:2005yw,Tanaka:2005ue,Barack:2009ux,Poisson:2011nh,Barack:2018yly,Barack:2018yvs,BHPT,BHPC}. 
The key goal of the EMRI modelling program is to generate waveforms 
for generic (eccentric and inclined) inspiraling orbits   
about astrophysical Kerr BHs, accounting for GSF effects.
In order to enable accurate extraction of EMRI parameters from a signal,
a GSF model must ultimately be accurate to the first subleading (`first \textit{post-adiabatic}') order in $\eta$~\cite{Detweiler:2005kq,Hinderer:2008dm,Miller:2020bft,vandeMeent:2020xgc,Pound:2021qin,Wardell:2021fyy}.
Attaining such accuracy will depend crucially upon calculations at leading order, which can be used to produce \textit{adiabatic} waveforms~\cite{Hughes:1999bq,Mino:2003yg,Hughes:2005qb,Mino:2005an,Mino:2006em,Mino:2007ft} as a baseline
in present-day data-analysis studies and for future science-adequate waveforms.

After decades of progress~\cite{Nakamura:1987zz,Finn:2000sy,Glampedakis:2002ya,Hughes:2001jr,Fujita:2004rb,Drasco:2005kz,Sundararajan:2007jg,Sundararajan:2008zm,Fujita:2009us,Harms:2013ib,Harms:2014dqa,Gralla:2016qfw,Burke:2019yek,Gourgoulhon:2019iyu,Fujita:2020zxe,Chua:2020stf}, 
adiabatic waveforms for generic Kerr orbits were obtained in 2021~\cite{Hughes:2021exa}.
Still, very few of these (mostly equatorial or spherical orbits) 
have been simulated so far, 
and the raw techniques used in that work are unsuitable in the provision 
of \textit{on-the-fly} waveforms for data-analysis studies. 
The main challenge is the high-dimensional parameter space of generic EMRIs. 
An adiabatic evolution must compute various inputs at fixed points in the parameter space 
in order to then drive the evolution through it. 
Populating the vast space sufficiently densely is a very computationally expensive task, 
and there are ongoing efforts to develop a cost-effective method 
to ease this burden~\cite{Barton:2008eb,Fujita:2020zxe,Chua:2020stf,Hughes:2021exa,Katz:2021yft}.

Meanwhile, there has been significant development of analytical techniques in GSF theory,
which invoke a `post-Newtonian' (PN) framework of BH perturbation 
theory~\cite{Mino:1997bx,Sasaki:2003xr}. 
This PN-GSF approach is far less expensive than numerical GSF calculations, 
readily covering a large region of parameter space that would be difficult 
to populate numerically~\cite{Fujita:2014eta,Sago:2015rpa,Kavanagh:2016idg,Bini:2018qvd,Bini:2019lcd,Munna:2020civ}.
To date, PN-GSF results have informed EMRI waveform model indirectly 
through `effective one-body'~\cite{Yunes:2009ef,Yunes:2010zj,Xin:2018urr,Zhang:2021fgy,Albanesi:2021rby} 
and `kludge' models~\cite{Glampedakis:2002cb,Barack:2003fp,Gair:2005ih,Babak:2006uv,Sopuerta:2011te,Chua:2015mua,Chua:2017ujo}.  
Despite being computationally cheap and much desired for LISA studies, 
the PN-GSF framework has not yet been implemented for relativistic adiabatic waveforms 
due to its technical complexity.

We here report the first adiabatic EMRI waveform model for generic Kerr orbits 
based on the analytical PN-GSF approach.
Our waveform is a standalone, on-the-fly model within GSF theory, 
efficiently covering the weak-field region 
of EMRI parameter space; see Figs.~\ref{fig:cos_err} and~\ref{fig:rel_err_PN} below. 
Among other things, this allows us to consistently account for 
the transient GSF resonances in the waveform~\cite{Tanaka:2005ue,Mino:2005an,Apostolatos:2009vu,Flanagan:2010cd,Brink:2013nna,Brink:2015roa}.
These resonances will play a crucial role in the detection and measurement of LISA EMRIs,  
but they have remained out of reach in high-cost relativistic evolutions   
and have so far only been treated with phenomenological models~\cite{Ruangsri:2013hra,Berry:2016bit,Speri:2021psr}. 

Furthermore, individual components of our model can be immediately combined with fits 
to numerical GSF data 
and environmental effects~\cite{Bonga:2019ycj,Gupta:2021cno,Coogan:2021uqv,Barsanti:2022ana} 
to construct both highly accurate and efficient
waveform models~\cite{Kludge_Suite,FEW} that will be employed 
in ongoing LISA preparatory studies (e.g.,~\cite{Chua:2021aah}), 
as well as in eventual production-level code for LISA's scientific analysis~\cite{LDC}.
They can be also used as reference points to inform the development of a universal model of binaries 
across all mass ratios~\cite{Pan:2010hz,Pratten:2020fqn,Rifat:2019ltp,Islam:2022laz}.
We expect that our PN-GSF approach will greatly improve the extensiveness 
and efficiency of EMRI modelling for LISA, 
in much the same way that analytical relativistic approaches 
have continued to advance modelling and data-analysis studies 
for LIGO, Virgo, and KAGRA~\cite{Ajith:2007kx,Aylott:2009ya,Ajith:2012az,Hinder:2013oqa}. 

Below we describe our adiabatic waveforms and assess their domain of validity. 
Throughout we set $G = c = 1$ and use $(t,\,r,\,\theta,\,\varphi)$ 
for Boyer-Lindquist coordinates. 

{\it{Snapshot waveforms from geodesic trajectories}}.---To set the stage, 
we first summarize EMRI snapshot waveforms, where the small body's motion 
is strictly geodesic~\cite{Drasco:2005kz,Sago:2005gd,Fujita:2009us}. 
$M$ and $a$ denote the mass and spin parameters of the Kerr BH;
$q \equiv a / M $, its dimensionless spin magnitude; 
and $\mu \ll M$, the mass of the small body (so $\mu/M = \eta$). 
Our convention is that $q>0$ ($q<0$) represents prograde (retrograde) orbits.

A bound Kerr geodesic is confined to a toroidal region 
given by $r_{\mathrm {min}} \leq r \leq r_{\mathrm {max}}$ and 
$\theta_{\rm min}\leq\theta\leq\pi-\theta_{\rm min}$.  
The orbit is generically triperiodic, uniquely described by three constants of motion $I_{A} \equiv \{p,\,e,\,\iota \}$~\cite{Cutler:1994pb,Hughes:1999bq} and three orbital phases $\Phi^A=\{\Phi^r,\Phi^\theta,\Phi^\varphi\}$ with constant frequencies $\rmd\Phi^A/\rmd t = \Omega^A(I_B)$~\cite{Schmidt:2002qk}. The three constant orbital parameters are the 
semi-latus rectum $ M p \equiv 
(2 r_{\mathrm {max}}\,r_{\mathrm {min}}) / ( r_{\mathrm {max}} + r_{\mathrm {min}} )$, 
orbital eccentricity $e \equiv 
({r_{\mathrm {max}} - r_{\mathrm {min}}}) / ( {r_{\mathrm {max}} + r_{\mathrm {min}}})$, 
and inclination angle $\tan \iota \equiv {\sqrt{\hat C}} / {{\hat L}}$, 
where ${\hat L}$ and ${\hat C}$ are the specific azimuthal angular momentum 
and Carter constant of the geodesic~\cite{Carter:1968rr}. The orbit's radial, polar, and azimuthal positions are $2\pi$-periodic in $\Phi^r$, $\Phi^\theta$, and $\Phi^\varphi$, respectively.

The gravitational radiation from these geodesics 
can be conveniently computed in the Teukolsky BH perturbation formalism, 
working with the linear perturbation of the Weyl scalar 
$\psi_4 = O(\eta)$~\cite{Teukolsky:1972my,Teukolsky:1973ha}; 
at infinity, the two GW polarizations $h_{+,\,\times}$ are 
simply given by 
$\psi_4(r \to \infty) = \frac{1}{2}\frac{\partial^2}{\partial t^2}
({h}_+ - i {h}_{\times})$. 
$\psi_4$ in the Fourier domain admits a full separation of variables by means of 
spin-weighted ($s=-2$) spheroidal harmonics ${{}_{-2} S_{\ell m}^{a \omega}}(\theta)$. The source geodesic's triperiodicity restricts the perturbation to the discrete frequency spectrum 
$
\omega_{mkn}\equiv
m\, \Omega^{\varphi} + k\, \Omega^{\theta} + n\, \Omega^{r}  
$ for integers $m,\,k,\,n$.
We may thus write $h_{+,\,\times}$ as a multipolar sum of ``voices" with those frequencies:
$
{h}_{+} - i {h}_{\times}
\propto 
r^{-1} \sum_{\lmkn} 
( {{Z}_{\lmkn}^{\infty}} {}_{-2} S_{\ell m}^{a {\omega}_{mkn}} / {{\omega}^2_{mkn}})
\, e^{- i \omega_{mkn} (t - r^*) + i m \varphi}
$  
with  
$\sum_{\lmkn} \equiv \sum_{\ell = 2}^{\infty} \sum_{m = -\ell}^{\ell} 
\sum_{k = -\infty}^{\infty} \sum_{n = -\infty}^{\infty}$ and  
the Kerr tortoise coordinate $r^*$~\cite{Teukolsky:1974yv}. 
Here, the dimensionless asymptotic amplitudes at infinity ${Z}_{\lmkn}^{\infty}(I_A)$ 
are obtained by solving the inhomogeneous radial Teukolsky equation 
mode by mode with a fixed geodesic source,  
enforcing outgoing (ingoing) boundary conditions at infinity (the horizon). 
The waveform phase $\omega_{mkn}\cdot(t-r^*)$ is a simple linear combination of the orbital phases, $\Phi_{mkn}(t)
\equiv m\, \Phi^{\varphi}(t) + k\, \Phi^{\theta}(t) + n\, \Phi^{r}(t)$, evaluated at the retarded time $u=t-r^*$.

{\it{Adiabatic waveforms from inspiral trajectories}}.---We now turn to adiabatic waveforms from orbits that slowly inspiral due to first-order [$O(\eta)$] 
GSF effects~\cite{Mino:1996nk,Quinn:1996am,Gralla:2008fg,Pound:2009sm}, 
specifically following the two-timescale framework developed in Refs.~\cite{Hinderer:2008dm,Miller:2020bft,Pound:2021qin}. 
In this framework, we introduce a `slow time' ${\tilde t} \equiv \eta t$. The trajectory and metric are treated as functions of 
slowly varying parameters ${I}_A(\tilde t) \equiv \{p({\tilde t}),\,e({\tilde t}),\,\iota({\tilde t}) \}$ and `fast time' phases $\Phi^A$ that evolve with slowly varying frequencies:
\begin{equation}\label{Phi-u}
\frac{\rmd\Phi^A}{\rmd t} =\Omega^A[{I}_B(\eta t)] 
\quad \text{or} \quad 
\frac{\rmd\Phi^A}{\rmd\tilde t} =\eta^{-1}\Omega^A[{I}_B(\tilde t)]\,, 
\end{equation}
where $\Omega^A({I}_B)$ is the same function of $I_B$ as for a geodesic. The dependence on $I_A$ captures the system's evolution on the radiation-reaction time scale $\sim M/\eta$, and the dependence on $\Phi^A$ capture the triperiodicity on the orbital time scale $\sim M$.

Adiabatic inspirals and waveforms in this framework can be heuristically understood as a slow-time evolution through the space of geodesic snapshots. The self-forced equations of motion and Teukolsky equations are split into slow- and fast-time equations. At each slow time step ${\tilde t}$, the leading-order fast-time equations are identical to the equations for a geodesic snapshot,
yielding the same Teukolsky amplitudes as a snapshot with the same parameters $I_A$. The adiabatic waveform can then be written in a form precisely analogous to the snapshot waveforms described above~\cite{Pound:2021qin,SM1}, 
\begin{equation}\label{IGW}
{h}_{+} - i {h}_{\times}
= 
-\frac{2\,\mu}{r} \sum_{\lmkn}\, 
\frac{{Z}_{\lmkn}^{\infty}}{{\omega}^2_{mkn}} \,
\frac{{}_{-2} S_{\ell m}^{a {\omega}_{mkn}} }{\sqrt{2 \pi}}\,
e^{-i {\Phi}_{mkn} 
+ i m \varphi } \,,
\end{equation}
where ${Z}_{\lmkn}^{\infty}$, ${\omega}^2_{mkn}$, and ${}_{-2} S_{\ell m}^{a {\omega}_{mkn}}$ are all geodesic functions of $I_A$, and the snapshot phase $\omega_{mkn}\cdot (t - r^*)$ 
is now replaced by the adiabatic phase 
$\Phi_{mkn} ({\tilde u})$ [satisfying Eq.~\eqref{Phi-u}]
evaluated at the slow retarded time $\tilde u \equiv \eta\,(t - r^*)$.

The evolution of $I_A$ can be derived from the self-forced equation of motion for the Boyer-Lindquist coordinate trajectory $z^\mu$. If we define {\it osculating} parameters $I^{\rm osc}_A$ by the condition that $z^\mu(I^{\rm osc}_A,\Phi^A)$ and $\rmd z^\mu/\rmd t(I^{\rm osc}_A,\Phi^A)$ satisfy the geodesic relationships between $\{z^\mu,\rmd z^\mu/\rmd t\}$ and $\{I_A,\Phi^A\}$, then we can straightforwardly derive an equation of the form $\rmd I^{\rm osc}_A/\rmd t=G_A\sim \eta$~\cite{Pound:2021qin}. If we write $I^{\rm osc}_A$ and $G_A$ as Fourier series of the form
$G_A =\sum_{{\mathsf k}_{r} {\mathsf k}_{\theta}} 
G_{A}^{{\mathsf k}_{r} {\mathsf k}_{\theta}}(I_B) \,
\exp{\{ i ( {\mathsf k}_{r}\, \Phi^{r} + {\mathsf k}_{\theta}\,\Phi^{\theta} ) \} }$, 
then at leading order the slowly varying $I_A$ is the stationary, $00$ mode of $I^{\rm osc}_A$, and its driving force is the $00$ mode of $G_A$. 
$G^{00}_A$ involves only the dissipative piece 
of $G_A$~\cite{Mino:2003yg,Mino:2005an,Tanaka:2005ue,Mino:2006em,Hinderer:2008dm} , 
which allows us to express it in terms of the asymptotic flux of radiation~\cite{Sago:2005gd}
in the convenient `flux-balance' form~\cite{Sago:2005fn,Drasco:2005is,Isoyama:2018sib}:  
\begin{equation}\label{dIdt}
\frac{\rmd {I}_{A}}{\rmd {\tilde t}} 
= 
- M\, 
\sum_{\lmkn}\, 
\frac{({\beta}_{mkn})_{A}}{4 \pi {\omega}^3_{m k n}}
\left\{
|{Z}_{\ell m k n}^{\infty} |^2
+
{\alpha}_{\ell m k n} \,
|{Z}_{\ell m k n}^{\mathrm {H}} |^2 
\right\}\,, 
\end{equation}
where ${Z}_{\ell m k n}^{\mathrm {H}}(I_A)$ is the Teukolsky amplitude 
of $\psi_4$ at the horizon, and ${\alpha}_{\lmkn}$ and $({\beta}_{mkn})_{A}$ are certain functions 
of ${I}_{A}$~\cite{Sago:2015rpa}. 
The adiabatic evolution is then given by Eqs.~\eqref{Phi-u}--\eqref{dIdt}.

{\it{Transient resonances}}.---However, the adiabatic-evolution scheme described above   
has to be altered if, at some instant ${\tilde t} = {\tilde t}_{\rm res}$, 
the slowly evolving orbital frequencies satisfy 
\begin{equation}\label{res-Omega}
\omega^{\rm res} ({\tilde t}_{\rm res}) 
\equiv 
{\beta^{r}} {\Omega^r}({\tilde t}_{\rm res}) - {\beta^{\theta}}{\Omega^{\theta}}({\tilde t}_{\rm res})  = 0\,, 
\end{equation}
with a pair of nonzero coprime integers $(\beta^{r},\,\beta^{\theta})$.  
This condition leads to a transient resonance of GSF effects that occurs 
for generic EMRIs~\cite{Tanaka:2005ue,Mino:2005an,Apostolatos:2009vu,Flanagan:2010cd,Brink:2013nna,Brink:2015roa}.
At the resonance, otherwise-oscillatory modes of $G_A$ 
with ${\mathsf k}_{r} : {\mathsf k}_{\theta} = {\beta}^{r} : -\beta^{\theta}$ become stationary,
and the flux-balance formulas~\eqref{dIdt} are then enhanced (or diminished) 
according to~\cite{Gair:2011mr,Grossman:2011im,Flanagan:2012kg,Isoyama:2013yor,vandeMeent:2013sza,Mihaylov:2017qwn,Isoyama:2018sib}
\begin{equation}\label{dIdt-res}
\frac{\rmd {I}_{A}^{{\rm res}}}{\rmd t} 
\equiv 
\left.
\sum_{s \neq 0}
G_A^{{\rm res}, s}
\,e^{i\,s\,\Phi^{\rm res}}
\right|_{ {\tilde t} = {\tilde t}_{\rm res} }\,.
\end{equation}
Here, $G_A^{{\rm res}, s}\equiv G^{s\beta_r,-s\beta_\theta}_A$ is the $s$th resonant mode of $G_A$, and the phase $\Phi^{\rm res} \equiv  \beta^{r} \, \Phi^{r} - \beta^{\theta}\, \Phi^{\theta}$ 
becomes stationary at resonance. 

Dissipation drives the orbit through the resonance, with a resonance-crossing time~\cite{Ruangsri:2013hra,Berry:2016bit}
\begin{equation}\label{t-res}
{\tilde T}^{{\rm res},\,s} 
\equiv
\sqrt{\frac{2 \pi} {|s\, {\dot \omega}^{{\rm res}}|}} 
\sim M\, \eta^{-1/2}\,,
\end{equation}
where 
${\dot \omega}^{{\rm res}} \equiv 
{\rmd {\omega}^{\rm res} / {\rmd \tilde t}}$
evaluated at the resonance.
$\tilde T^{{\rm res},\, s}$ is longer than the orbital time scale $\sim M$, but much shorter than 
the radiation-reaction time scale $\sim  M / \eta$. 
On the radiation-reaction time, the resonance crossing is effectively instantaneous, and the correction~\eqref{dIdt-res} 
causes a sudden jump $\delta I_A^{\rm res}$ at ${\tilde t} = {\tilde t}_{\rm res}$.  
Formally expanding the phase $\Phi^{\rm res}(t)$ around the resonance 
as $\Phi^{\rm res}(t_{\rm res}) + \frac{1}{2}{\dot \omega}^{{\rm res}} (t - t_{\rm res})^2 + O(|t-t_{\rm res}|^3)$
with the aid of Eqs.~\eqref{Phi-u} and~\eqref{res-Omega}, 
and integrating Eq.~\eqref{dIdt-res} 
under the stationary phase approximation, 
we find (see, e.g., Refs.~\cite{Barack:2018yvs,Pound:2021qin}) 
\begin{equation}\label{DeltaI}
\delta I_A^{\rm res} 
=
\sum_{s \neq 0}\left.\, 
{\tilde T}^{{\rm res},\,s} \, {G_{A}^{{\rm res}, s}}\, 
e^{{\rm sgn} \left( s\, {\dot \omega}^{{\rm res}} \right) \frac{i\pi}{4} 
+ i\,s\,\Phi^{\rm res}}\,
\right|_{ {\tilde t} = {\tilde t}_{\rm res} }\sim \eta^{1/2}\,.
\end{equation}
This induces corresponding abrupt frequency jumps $\delta\Omega^A\sim \eta^{1/2}/M$, 
resulting in large cumulative phase shifts $\delta\Phi^A\sim \eta^{-1/2}$ a radiation-reaction time 
after the resonance-crossing, which will deteriorate our ability to measure EMRIs.

If we only have access to the leading-order dissipative GSF, 
we cannot calculate the exact $O(\eta^{1/2})$ jumps for two reasons. 
First, the size of $\delta I_A^{\rm res}$ in Eq.~\eqref{DeltaI}
sensitively depends on the phase $\Phi^{\rm res}$. 
Calculating the jump therefore requires knowing the phase through 
first post-adiabatic order in the evolution preceding the resonance~\cite{Lukes-Gerakopoulos:2021ybx}.
Second, the conservative GSF directly contributes to 
$G_{A}^{{\rm res}, s}$~\cite{Fujita:2016igj,Isoyama:2018sib}. 
Accounting for these effects is beyond the current state of the art.

Nevertheless, Eq.~\eqref{DeltaI} correctly determines the jump that is internally consistent with an adiabatic phase evolution. Similarly, discarding the conservative contribution to Eq.~\eqref{DeltaI} yields the correct jump associated with the dissipative GSF. 
That dissipative piece of $G_A^{{\rm res}, s}$ can be constructed directly 
from the Teukolsky amplitudes as~\cite{SM2}
\begin{align}\label{eq:GSF-res}
& 
G_A^{{\rm res}, s}(I_B)
= 
- M\, \sum_{\lmkn}\, 
\sum_{\substack {k' = k - s \beta^{\theta} \\ n' = n + s \beta^{r}}}
\frac{({b}_{m k n})_{A} + {k'}\, ({c}_{m k n})_{A}}{4 \pi {\omega}^3_{m k n}}  \cr 
& \times 
\left\{
{\rm Re} \left(
{Z}_{\ell m k n}^{\infty}\, \overline {{Z}_{\ell m k' n'}^{\infty}}
\right)
+
{\alpha}_{\ell m k n} \,
{\rm Re} \left(
{Z}_{\ell m k n}^{\rm H}\, \overline {{Z}_{\ell m k' n'}^{\rm H}}
\right)
\right\}\,,
\end{align}
where the overline denotes complex conjugation,
and $\{({b}_{mkn})_A,\,(c_{mkn})_{A}\}$ are certain functions of ${I}_{A}$. 
Because both conservative and dissipative contributions are comparable 
(at least in a scalar-field toy model~\cite{Nasipak:2021qfu}), 
we expect the jumps calculated in this way should qualitatively capture 
the impact of a resonance crossing, at the order-of-magnitude level. 

{\it{Techniques}}.---The end-to-end implementation of Eqs.~\eqref{Phi-u}, \eqref{IGW}, \eqref{dIdt}, 
and~\eqref{DeltaI} to generate adiabatic waveforms needs three main inputs 
across the full parameter space of Kerr spin $q$ and orbital parameters $I_A$: 
the frequencies ${\Omega}^A(I_B)$, the spheroidal harmonics ${}_{-2} S_{\ell m}^{a {\omega}_{mkn}}$, 
and the Teukolsky amplitudes ${Z}_{\lmkn}^{\infty,\,{\mathrm {H}}}$. 
The expression for ${\Omega}^A(I_B)$ is given 
in closed analytical form~\cite{Schmidt:2002qk,Fujita:2009bp}. 
We can also analytically compute ${}_{-2} S_{\ell m}^{a {\omega}_{mkn}}$ and 
${Z}_{\lmkn}^{\infty,\,{\mathrm {H}}}$, 
building on the semi-analytical method 
of solving the Teukolsky equation~\cite{1977JMP....18.1849F,Mano:1996vt,Mano:1996gn} 
in a small-frequency expansion. 
This calculation is performed with the analytical module 
of the Black Hole Perturbation Club (BHPC) code~\cite{BHPC} 
developed in Refs.~\cite{Shibata:1994jx,Tagoshi:1995sh,Tagoshi:1996gh,Tanaka:1996lfd,Tagoshi:1997jy,Sago:2005fn,Ganz:2007rf,Fujita:2011zk,Fujita:2012cm,Fujita:2014eta,Sago:2015rpa}. 
It assumes the `spheroidicity' $a \omega$, `velocity' ${v} \equiv \sqrt{1/{p}}$, 
and eccentricity ${e}$ are much smaller than unity 
but allows arbitrary inclination ${\iota}$ and spin $|q| < 1$~\cite{Ganz:2007rf}.
We obtain ${}_{-2} S_{\ell m}^{a {\omega}_{mkn}}$ up to $(a \omega_{mkn})^4$ (so $\sim v^{12}$), 
and ${Z}_{\lmkn}^{\infty,\,{\mathrm {H}}}$ through $v^{10}$ and $e^{10}$ 
beyond their leading order `Newtonian-circular' terms~\cite{BHPC-Data}. 
This `$5$PN-$e^{10}$' calculation includes harmonic modes 
of $2 \leq \ell \leq 12$ with $|m| \leq \ell$, $|m + k| \leq 12$ and $|n| \leq 10$, 
which gives $\approx 33,000$ nontrivial modes in total after exploiting 
mode symmetries~\cite{Drasco:2005kz,Fujita:2009us,vandeMeent:2017bcc,Nasipak:2019hxh}.

With the analytical inputs in hand, we numerically evolve 
the system of GW phases~\eqref{Phi-u}, polarizations~\eqref{IGW},  
and orbital parameters~\eqref{dIdt} in the slow time $\tilde t$; 
this numerical evolution (mostly) avoids the $5$PN-$e^{10}$ expansion of ${\Omega}^A(I_B)$, 
which would severely limit waveform accuracy. 
The evolution starts with given initial parameters $I_A (0) \equiv I_A({\tilde t} = 0)$ 
and phases $\Phi^A(0) \equiv \Phi^A(\tilde t=0)$  
and halts when it satisfies the resonance condition~\eqref{res-Omega}, 
at which time the jump $\delta I_A^{\rm res}$ is calculated from Eq.~\eqref{DeltaI}.   
We then resume the evolution from ${\tilde t} = {\tilde t}_{\rm res}$ 
with the shifted orbital parameters $I_A({\tilde t_{\rm res}}) + \delta I_A^{\rm res}$. 
These procedures are repeated until the evolution reaches a chosen termination time. 
The nonresonant part of the evolution is implemented for CPUs, 
and it is competitive with the numerical kludge code~\cite{Babak:2006uv}, 
which is the `fastest' (semi-relativistic) on-the-fly waveform so far.

{\it{Domain of validity}}.---Before presenting our adiabatic waveform, 
we compare against an `exact' numerical adiabatic data set 
to assess the accuracy of our $5$PN-$e^{10}$ Teukolsky amplitudes and fluxes.  

In doing so, we also employ the numerical module 
of the BHPC code~\cite{BHPC,Fujita:2004rb,Fujita:2009uz,Fujita:2009us,Fujita:2020zxe} 
to compute ${}_{-2} S_{\ell m}^{a {\omega}_{mkn}}$ 
for viewing angles $0^{\circ} \leq \theta \leq 90^{\circ}$, 
and ${Z}_{\lmkn}^{\infty,\,{\mathrm {H}}}$ 
for $q = \{0,\,\pm 0.3,\,\pm 0.5,\,\pm 0.7,\,\pm 0.9\}$ 
and 
$\iota \sim \{0^{\circ},\,20^{\circ},\,40^{\circ},\,60^{\circ},\,80^{\circ}\}$  
on a grid in $(p,\,e)$, where $6 \lesssim p \lesssim 30$ 
and $0.01 \lesssim e \lesssim 0.4$.
The harmonic content of the numerical data is dynamically determined 
by the condition that the corresponding fluxes 
${\cal F}_A^{\rm num} \equiv {\rmd {I}_{A}}/{\rmd {\tilde t}}$ 
obtained from Eq.~\eqref{dIdt} have fractional accuracy $ \lesssim 10^{-5}$.
We then use two figures of merit for the $5$PN-$e^{10}$ analytical results: 
(i) the `mode-distribution error' of vectorized amplitudes $H \equiv {\rm{vec}} 
\left({{Z}_{\lmkn}^{\infty}}\, {}_{-2} S_{\ell m}^{a {\omega}_{mkn}}/{{\omega}^2_{mkn}}\right)$ defined 
by $|1 - \Re{(H^{{\rm ana}},\,H^{\rm num})} / (|H^{\rm ana}|\,|H^{\rm num}|)|$, 
where the inner product $(\cdot\,,\cdot)$ and its associated norm $|\cdot|$ 
are (implicitly) Hermitian~\cite{Chua:2020stf}, 
and (ii) the relative flux error of $I_A$ 
defined by $|1 - {\cal F}_A^{\rm ana} / {\cal F}_A^{\rm num}|$.

Figures~\ref{fig:cos_err} and~\ref{fig:rel_err_PN} show example results 
of the mode-distribution error and relative flux error, respectively~\cite{SM3}. 
In addition to the obvious limitation 
in the small-$p$ (hence large-$v$) and large-$e$ parameter regions, 
the errors are also larger with decreasing $\iota$ or $q$  
when the last stable geodesic orbit lies at $p \gtrsim 6.0$~\cite{Stein:2019buj}; 
the location of this strong-field orbit  
is difficult to capture with a small-$(v,\,e)$ expansion.  
In Fig.~\ref{fig:rel_err_PN}, we also test Teukolsky-fitted fluxes 
used in kludge models~\cite{Glampedakis:2002cb,Gair:2005ih,Babak:2006uv}.  
In general, the $5$PN-$e^{10}$ fluxes are more accurate than the fitted 
fluxes for $e \gtrsim 0.2$, independent of $q$ and $\iota \neq 0$.

\begin{figure}[tbp]
	\includegraphics[angle=0,width=\columnwidth]{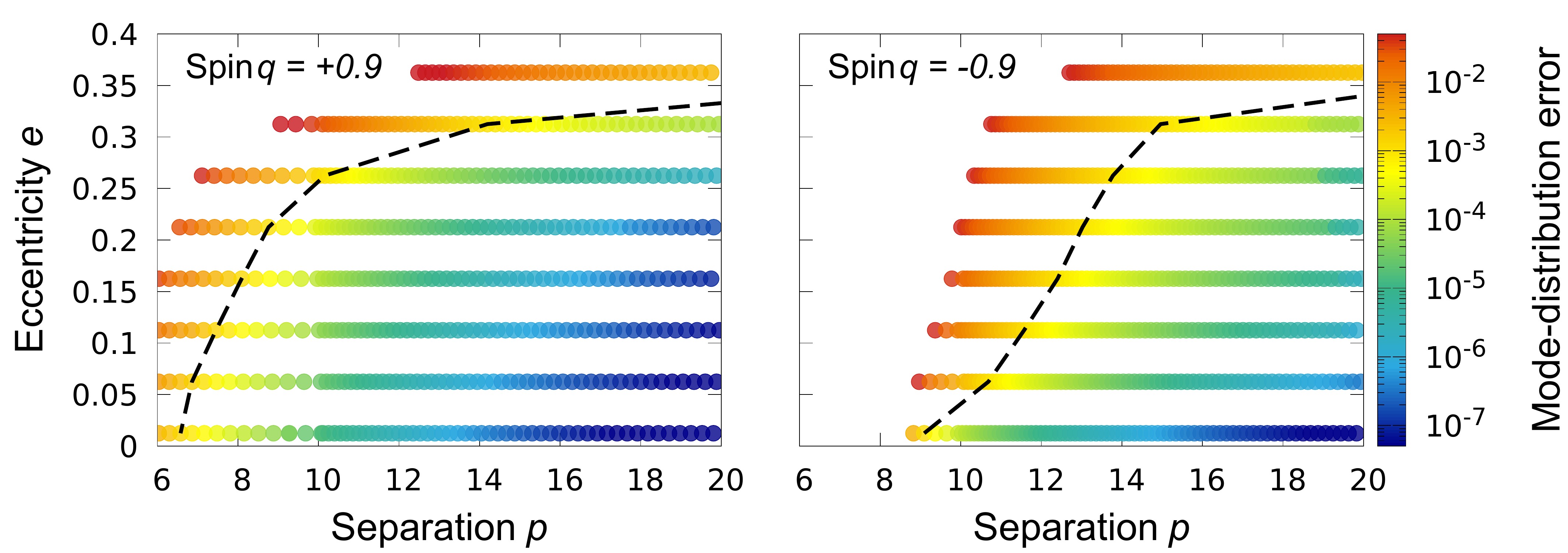}
	\caption{Mode distribution error of $5$PN-$e^{10}$ amplitudes 
	relative to numerical data.
	We show the best case of $q = +0.9$, $\iota \approx 80^{\circ}$ (\textit{Left})
	and the worst case of $q = -0.9$, $\iota \approx 20^{\circ}$ (\textit{Right}); 
	for the latter, the last stable orbit always lies at $p > 8.5$.
    The dashed curves indicate an error of $\approx 1.0 \times 10^{-3}$, 
    and at $p = 20$ they cross $e \approx 0.33$ ($e \approx 0.35$)
    in the left (right) panel. 
	The plots are for $(\theta,\,\varphi) = (45^{\circ},\,0^{\circ})$
	but do not depend strongly on the viewing angle 
    because our expansion of ${}_{-2} S_{\ell m}^{a {\omega}_{mkn}}$  
    accounts for higher $(a \omega_{mkn})^4$-corrections, valid up to $7$PN. 
	}
	\label{fig:cos_err}
\end{figure}

\begin{figure}[tbp]
    \includegraphics[angle=0,width=\columnwidth]{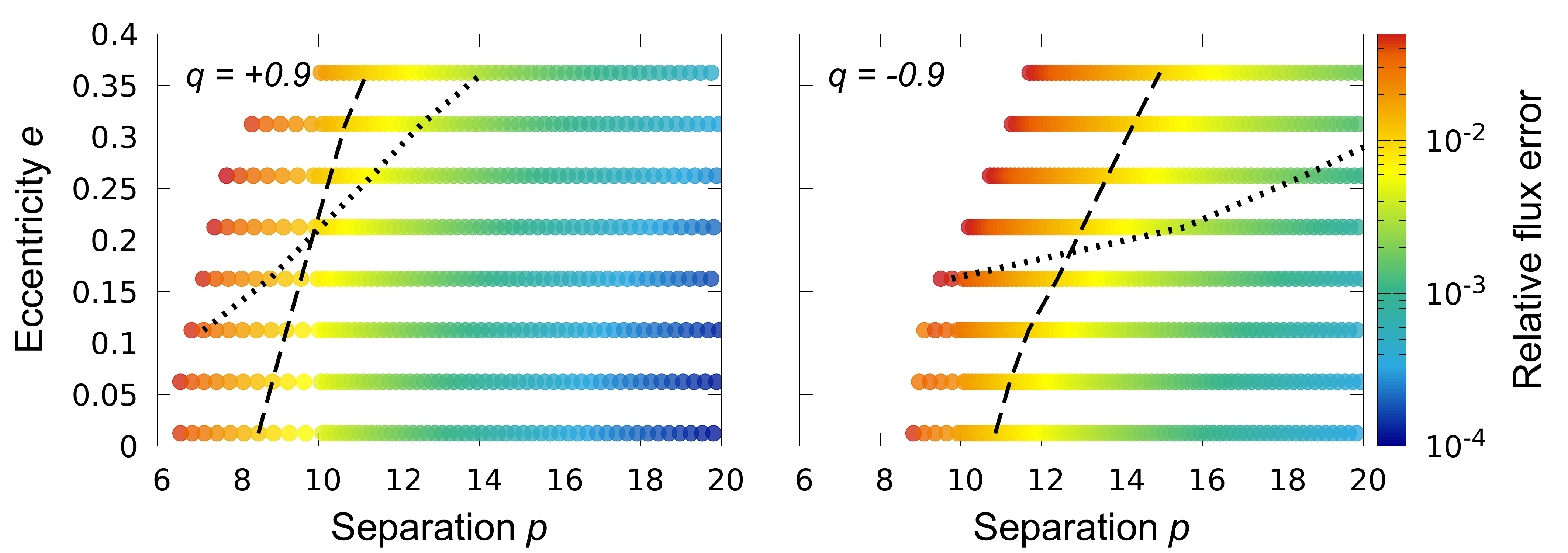}
	\caption{Error of $5$PN-$e^{10}$ fluxes ${\cal F}_{p} (\equiv {\rmd p} / {\rmd {\tilde t}})$
	relative to numerical flux data,
	with the same parameters as those in Fig.~\ref{fig:cos_err}. 
	The dashed curves indicate an error of $\approx 1.0 \times 10^{-2}$, 
	while the dotted curves show, for reference, 
	the same level of error for Teukolsky-fitted fluxes~\cite{Glampedakis:2002cb,Gair:2005ih,Babak:2006uv}.
	}
	\label{fig:rel_err_PN}
\end{figure}

For parameter inference of LISA-type EMRIs, a mode distribution error 
of $ \lesssim 1.0 \times 10^{-3}$ is adequate~\cite{Chua:2020stf,Katz:2021yft}. 
Inference requirements are far more stringent for the fluxes, but these will not be attained even with `exact' adiabatic models.
Comparisons to numerical adiabatic evolutions of equatorial orbits~\cite{Fujita:2020zxe} suggest that relative flux errors $\lesssim 10^{-2}$ (of $dp/d{\tilde t}$) will suffice for our waveform 
to maintain phase coherence with adiabatic LISA-EMRI waveforms over several months; 
this will also be the level of agreement between adiabatic and post-adiabatic models~\cite{SM4}.
We therefore estimate the domain of validity of the $5$PN-$e^{10}$ results    
as $6 \lesssim p \lesssim 20$ and $0 \lesssim e \lesssim 0.3$ across $|q| \leq 0.9$ and $0^{\circ} \leq \iota \leq 80^{\circ}$, 
excluding the parameter region near the last stable orbit.

{\it{Sample results}}.---As a concrete example, we present a $5$PN-$e^{10}$ adiabatic waveform 
with masses and Kerr spin $(\mu,\,M,\,q) = (10 M_{\odot},\,10^6M_{\odot},\,0.9)$, 
initial orbital parameters $(p(0),\,e(0),\,\iota(0)) = (9.6,\,0.21,\,80^{\circ})$, 
and initial phases $\Phi^A(0) = 0$. 
We evolve the EMRI over $\approx 4$ months, starting from the initial apastron 
$(r(0),\, \theta(0),\,\varphi(0)) = [M p(0) / (1 - e(0)),\,\pi / 2,\,0]$ 
with $d \theta(0) / dt < 0$, 
and ending with final orbital parameter values 
$(p_{\rm f},\,e_{\rm f},\,\iota_{\rm f}) \approx (9.23,\,0.197,\,80.1^{\circ})$. 
During that time the inspiral
passes through one strong, $3$:$2$ resonance; 
all other resonances that it encounters are suppressed by additional powers of $v$ and $e$ 
and contribute negligible jumps according to Eq.~\eqref{DeltaI}.

\begin{figure*}[tbp]
	\includegraphics[angle=0,width=2.06\columnwidth]{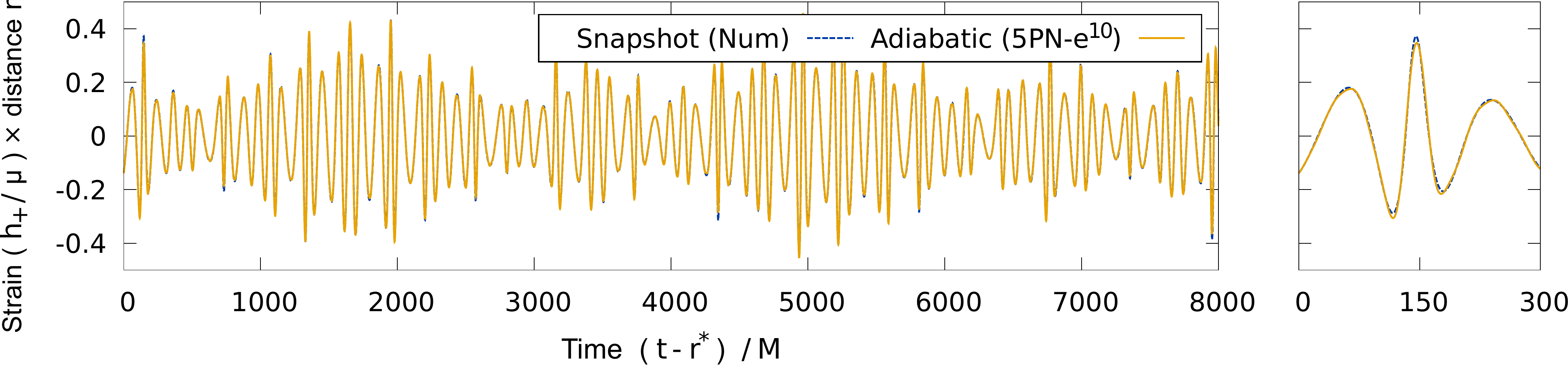}
	\caption{
	\textit{Left}: $5$PN-$e^{10}$ adiabatic waveform of a sample EMRI  
	with masses $(M,\,\mu) = (10^6 M_{\odot},\,10 M_{\odot})$ and spin $q = 0.9$, 
	starting at $(p_0,\,e_0,\,\iota_0) = (9.6,\,0.21,\,80^{\circ})$. 
	We plot the first $\approx 11$ hours ($M \approx 5$~sec) 
	of the evolving waveform at the viewing angle $(\theta,\,\varphi) = (45^{\circ},\,0^{\circ})$. 
	The rich structure of the waveform is due to the beating of voice sets $\omega_{mkn}$, which
    encode, for example, periastron precession $(\propto \Omega^{\varphi} - \Omega^{r})$ 
    and Lense-Thirring precession $(\propto \Omega^{\varphi} - \Omega^{\theta})$.
	The dashed curve is the reference snapshot waveform 
	from the fixed geodesic orbit with the same initial frequencies and phases at $t = 0$; this was generated 
	using the BHPC's numerical Teukolsky code~\cite{Fujita:2009us}. 
	\textit{Right}: the first $\approx 25$ minutes. 
	} 
	\label{fig:strain}
\end{figure*}

Figure~\ref{fig:strain} shows the first $\approx 11$ hours of 
the $5$PN-$e^{10}$ adiabatic strain $h_{+}$.  
For reference, we also plot the snapshot strain 
from the fixed geodesic source (i.e., without GSF effects) 
with the same initial frequencies, generated by the BHPC's numerical Teukolsky 
solver~\cite{Fujita:2009us}; 
this serves as a benchmark for the $5$PN-$e^{10}$ waveform so long as 
the viewing time is much shorter than the dephasing time
$ \sim M / (\sqrt{\eta} \, \Omega^A)$~\cite{Detweiler:2005kq,Barack:2018yvs},
after which the inspiral orbit becomes $\sim 1$ cycle out of phase with the geodesic orbit. 
The dephasing time of this sample EMRI is $\sim 2$ days.
As the figure shows, the $5$PN-$e^{10}$ model faithfully approximates the numerical snapshot, 
a consequence of the small mode distribution error $\approx 5.0 \times 10^{-4}$ 
in the $5$PN-$e^{10}$ amplitudes.

\begin{figure}[tbp]
    \includegraphics[angle=0,width=\columnwidth]{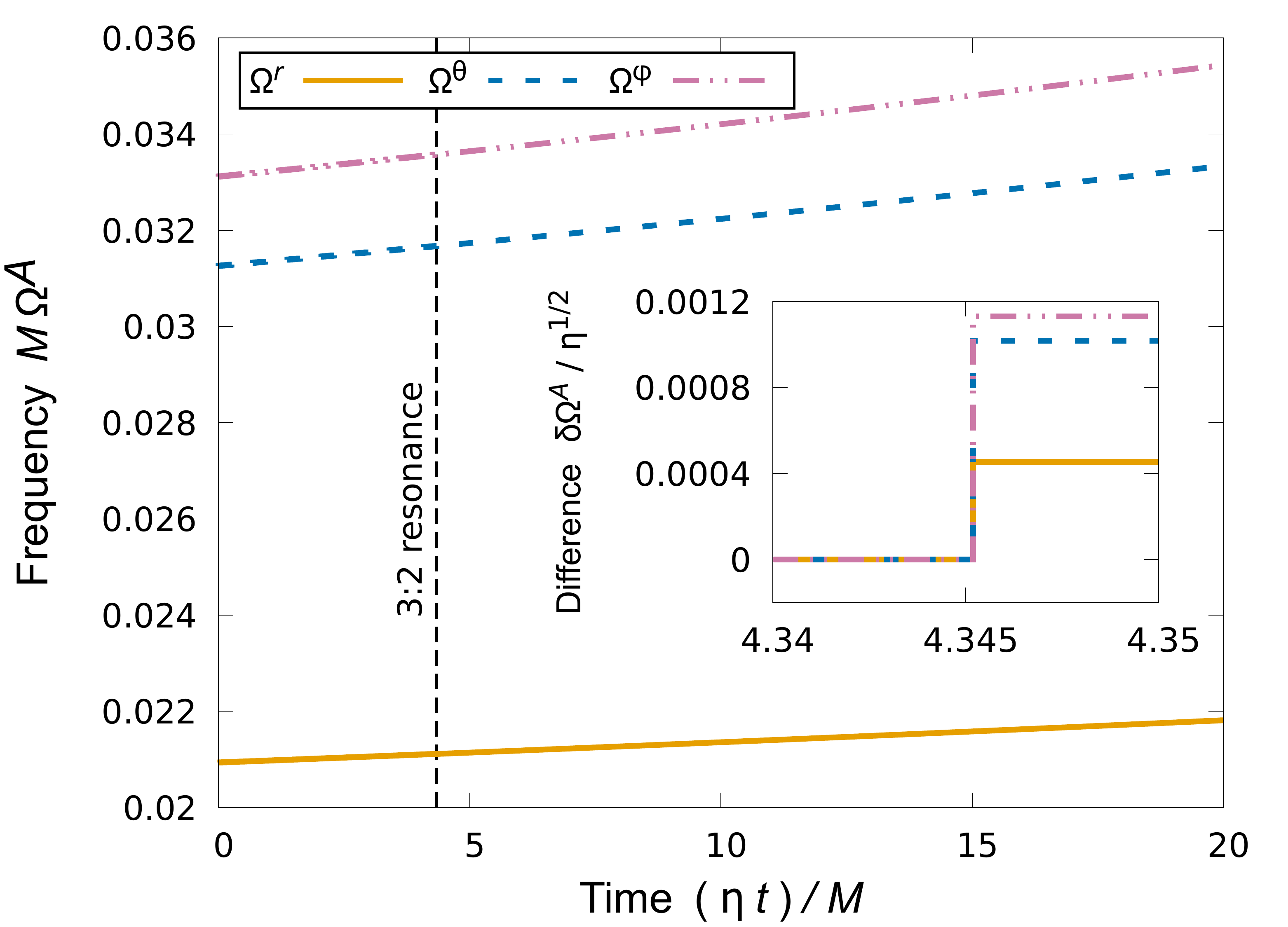}
	\caption{Slowly evolving orbital frequencies 
	$\Omega^A \equiv \{\Omega^{r},\,\Omega^{\theta},\,\Omega^{\varphi}\}$ 
	for the sample EMRI waveform in Fig.~\ref{fig:strain} 
	as a function of the slow time ${\tilde t} = \eta\,t$.
	The adiabatic evolution lasts for $\approx 4\text{ months}$ $({\tilde t} = 20.0\,M)$. 
	The vertical dashed line marks the $3$:$2$ resonance 
    at ${\tilde t}_{\rm res} \approx 4.3452\,M$ 
    with frequencies $M\,\Omega^r \approx 2.11\times 10^{-2}$ 
    and $M\,\Omega^{\theta} \approx 3.17\times 10^{-2}$. 
	The inset enlarges the region near the resonance, 
	showing the difference $\delta \Omega^A := M(\Omega^{A}_{\rm non-res} - \Omega^{A}_{\rm res})$
	between the evolution with and without resonance effects. 
    At the resonance, there are order-$\sqrt{\eta}$ discontinuous frequency jumps. 
	}
	\label{fig:time-freq}
\end{figure}

The slow evolution of the adiabatic waveform is more visible 
in the time-frequency plot~\ref{fig:time-freq}. 
The $3$:$2$ resonance occurs at ${\tilde t}_{\rm res} \approx 4.3452\,M$, 
where there are abrupt frequency jumps 
$(\delta \Omega^{r} / {\sqrt {\eta}},\, \delta \Omega^{\theta} / {\sqrt {\eta}},\,
     \delta \Omega^{\varphi}  / {\sqrt {\eta}}) 
     \approx (4.54 \times 10^{-4},\, 1.02 \times 10^{-3},\, 1.13 \times 10^{-3})$, 
corresponding to the jumps  
$(\delta p^{\rm res} / {\sqrt {\eta}},\, \delta e^{\rm res} / {\sqrt {\eta}},\,
\delta {\iota}^{\rm res}  / {\sqrt {\eta}}) 
\approx (1.80 \times 10^{-2},\, 8.66 \times 10^{-3},\, 2.17^{\circ} \times 10^{-2})$
estimated from Eq.~\eqref{DeltaI}.  
Although the frequency jumps are small $(\propto \sqrt{\eta})$,  
they lead to large cumulative phase shifts $(\delta \Phi^{\rm res}_r,\,\delta \Phi^{\rm res}_{\theta},\,\delta \Phi^{\rm res}_{\varphi}) 
\approx (2.38,\,5.49,\,6.13)$ by the termination time ${\tilde t} = 20.0\,M$. 
Such shifts are dramatic compared to LISA's EMRI phase resolution $\sim  0.1$~rad~\cite{Lindblom:2008cm,Bonga:2019ycj,Gupta:2021cno}, 
reconfirming the importance of GSF resonances 
for EMRI measurements~\cite{Ruangsri:2013hra,Berry:2016bit,Speri:2021psr}.

{\it{Concluding remarks}}.---Our PN-GSF adiabatic model 
represents the first user-ready, relativistic description of EMRI waveforms 
in the astrophysical scenario of generic Kerr orbits,  
including an approximate treatment of GSF resonances.
It can be used to generate on-the-fly waveforms over the whole weak-field, small-eccentricity region of the EMRI parameter space, with arbitrary orbital inclination and Kerr spin, thus opening a new front in ongoing EMRI modelling and data analysis efforts~\cite{Kawamura:2020pcg,Chua:2021aah}. 

In the near term, we will improve our $5$PN-$e^{10}$ analytical calculations to cover more of the EMRI parameter space~\cite{Fujita:2012cm,Fujita:2014eta,Shah:2014tka,Munna:2019fjz,Munna:2020juq,Munna:2020iju,Munna:2020som}. 
We will further accelerate our model towards EMRI data-analysis, 
using the efficiency-oriented {\sc FastEMRIWaveforms} framework~\cite{FEW}, 
which will enable a highly parallelized implementation 
with graphics processing units~\cite{Chua:2020stf,Katz:2021yft}. 
Ultimately, we will work on refining the adiabatic model 
by combining analytical PN-GSF results with numerical GSF data~\cite{Warburton:2011fk,Osburn:2015duj,vandeMeent:2017bcc,vandeMeent:2018rms,Pound:2019lzj,Warburton:2021kwk,Fujita:2020zxe,Hughes:2021exa,McCart:2021upc,Warburton:2021kwk,Lynch:2021ogr,Mathews:2021rod,Skoupy:2022adh}, 
to accomplish a science-adequate, post-adiabatic waveform for LISA.  

Finally, it would be informative to compare our adiabatic evolution  
with small-mass-ratio results from PN theory~\cite{Will:2016pgm,Tucker:2021mvo} and fully nonlinear 
numerical-relativity simulations~\cite{Lewis:2016lgx,Fernando:2018mov,Lousto:2020tnb,Lousto:2022hoq}. 
This may further delineate the applicable region of GSF theory~\cite{Tiec:2014lba,vandeMeent:2020xgc}  
for generic binary BHs.

\begin{acknowledgments}
{\it{Acknowledgments}}.---We thank Wataru Hikida and Hideyuki Tagoshi 
for their direct contributions to an earlier version of this manuscript;  
Scott A. Hughes for helpful discussions on initial phases and comments on the Supplemental Material; 
Maarten van de Meent for providing independent numerical data 
to verify BHPC's Teukolsky results; 
Leor Barack and Chulmoon Yoo for valuable discussions and comments on the manuscript; 
and Katsuhiko Ganz, Chris Kavanagh, Koutarou Kyutoku, Yasushi Mino, Takashi Nakamura, 
Misao Sasaki, Masaru Shibata, and Niels Warburton for very helpful discussions.  
S.I. is especially grateful to Eric Poisson, Riccardo Sturani and Takahiro Tanaka 
for their continuous encouragement and insightful discussion 
about the (adiabatic) evolution scheme for EMRI dynamics. 
Finally, we thank all the past and present members of the annual Capra meetings 
with whom we have discussed the techniques and results presented here (over the past decades).

S.I. acknowledges support from STFC through grant no. ST/R00045X/1,  
the GWverse COST Action CA16104,``Black holes, gravitational waves and fundamental physic'',  
and additional financial support of Ministry of Education - MEC 
during his stay at IIP-Natal-Brazil.
A.J.K.C. acknowledges support from the NASA grants no. 18-LPS18-0027 and 20-LPS20-0005, 
and from the NSF grant no. PHY-2011968.
A.P. acknowledges the support of a Royal Society University Research Fellowship,
Research Grant for Research Fellows, Enhancement Awards, and Exchange Grant. 
This work was supported in part by JSPS/MEXT KAKENHI Grant 
no.\ JP16H02183 (R.F.), JP18H04583 (R.F.), JP21H01082 (R.F., H.N. and N.S.),
JP17H06358 (H.N. and N.S.), and JP21K03582 (H.N.).
\end{acknowledgments}


\bibliographystyle{apsrev4-2}
\bibliography{PNadiabatic}

\end{document}



\title{Supplemental material}

\maketitle


\setcounter{equation}{0}
\setcounter{figure}{0}
\setcounter{table}{0}
\setcounter{page}{1}
\makeatletter
\renewcommand{\theequation}{S\arabic{equation}}
\renewcommand{\thefigure}{S\arabic{figure}}
\renewcommand{\bibnumfmt}[1]{[S#1]}
\renewcommand{\citenumfont}[1]{S#1}

In this supplemental material, we provide further insight 
into our $5$PN-$e^{10}$ adiabatic model, 
including 
(i) a comparison with another adiabatic-evolution scheme 
due to Hughes et al.~\cite{Hughes:2021exa}, 
(ii) a sketch of the derivation of the evolution equations 
across resonances, 
(iii) additional plots of relative flux errors,   
and 
(iv) an estimate of the dephasing between adiabatic 
and first post-adiabatic model. 

Notations, conventions and symbols are all the same as those 
used in the body of the letter, unless stated otherwise.

\section{Initial phases and choice of angle variables}

Our adiabatic evolution differs in one detail from the evolutions of Hughes et al.~\cite{Hughes:2021exa}. The difference arises in our treatment of initial phases. Although we write the Teukolsky mode amplitudes as functions of the orbital parameters $I_A$, they are also implicitly functions of the initial phases. Here we show how this dependence affects the evolving waveform in the context of the two-timescale expansion.

The initial-phase dependence of the Teukolsky amplitudes is inherited directly from the initial-phase dependence of the modes of the point-particle stress-energy tensor $T^{\mu\nu}$. We can therefore restrict our attention to $T^{\mu\nu}$, which reads
\begin{align}\label{Tmunu}
T^{\mu\nu} &= \mu\,\frac{u^{\mu} u^{\nu}}{u^t \Sigma}\,
\delta[r - r_{\rm p}(t)]\,\delta[z - z_{\rm p}(t)]\,\delta[\varphi - \varphi_{\rm p}(t)], 
\end{align}
where $z\equiv\cos\theta$, $x^\mu_{\rm p}(t)$ denotes the particle's trajectory as a function of Boyer-Lindquist time $t$, $u^\mu\equiv dx^\mu_{\rm p} / d\tau = u^t dx^{\mu} / dt$, and $\Sigma\equiv r^2 + a^2 z^2$.

Whether we consider snapshots or the evolving orbit, the functions $r_{\rm p}(t)$, $z_{\rm p}(t)$, and $u^\mu(t)$ are periodic functions of the phase variables $\Phi^r(t)$ and $\Phi^\theta(t)$. The azimuthal angle $\varphi_{\rm p}(t)$ can be written as $\varphi_{\rm p}(t) = \Phi^\varphi(t) + \Delta \varphi(t)$, where $\Delta \varphi(t)$ is an oscillatory function of $\Phi^r(t)$ and $\Phi^\theta(t)$. It follows that we can write $T^{\mu\nu}$ as a discrete Fourier series in the phases, 
\begin{equation}\label{generic}
T^{\mu\nu} = \sum_{mkn}T^{\mu\nu}_{mkn}(r,\theta)e^{-i\Phi_{mkn}(t)+im\varphi},
\end{equation}
where we write $T^{\mu\nu}=T^{\mu\nu}(\Phi^A(t),r,\theta,\varphi)$ and define $T^{\mu\nu}_{mkn}\equiv \frac{1}{(2\pi)^3}\oint d^3\Phi\, T^{\mu\nu}(\Phi^A,r,\theta,\varphi)e^{i\Phi_{mkn}}$. The values of $T^{\mu\nu}_{mkn}$ obviously depend on how the phases relate to the coordinate position. For concreteness, choose the origin of the $\Phi^A$ phase space such that $\Phi^A=0$ at $r_{\rm p}=r_{\rm min}$, $\theta_{\rm p}=\theta_{\rm min}$ ($z_{\rm p}=z_{\rm max}$), and $\varphi_{\rm p}=0$ (where  these coordinate positions are defined for fixed parameters $I_A$); this corresponds to the choice in Pound \& Wardell~\cite{Pound:2021qin}, and the coefficients $T^{\mu\nu}_{mkn}$ are then given explicitly by Eq. (411) of that reference. Note that this is simply a choice of coordinates on the phase space rather than a statement about any particular orbit $x^\mu_{\rm p}(t)$ in that phase space. The orbit might not pass through the point $(r_{\rm min},\theta_{\rm min},0)$, and likewise it might never pass through the origin in the phase space.

In the case of a geodesic snapshot, the phases are given by $\Phi^A(t) = \Omega^A t + \Phi^A(0)$. We then have
\begin{equation}\label{geodesic}
T^{\mu\nu} = \sum_{mkn}T^{\mu\nu}_{mkn}(r,\theta)e^{-i\Phi_{mkn}(0)} e^{-i\omega_{mkn}t+im\varphi}.
\end{equation}
In typical applications, one solves field equations for coefficients of $e^{-i\omega_{mkn}t+im\varphi}$, with sources $\hat T^{\mu\nu}_{mkn} = T^{\mu\nu}_{mkn}e^{-i\Phi_{mkn}(0)}$. Without loss of generality, we can choose $t$ and $\varphi$ such that $t=0=\varphi_{\rm p}(0)$ at a periapsis passage, implying  $\Phi^\varphi(0)=\Phi^r(0)=0$. With these choices, $\hat T^{\mu\nu}_{mkn} = T^{\mu\nu}_{mkn}e^{-ik\Phi^\theta(0)}$. For a fiducial geodesic with $\theta_{\rm p}(0)=\theta_{\rm min}$, we also have $\Phi^\theta(0)=0$, implying that for this geodesic, $\hat T^{\mu\nu}_{mkn} = T^{\mu\nu}_{mkn}$. If one computes the Teukolsky mode amplitude $Z_{\lmkn}$ for this fiducial geodesic, one can then recover the $Z_{\lmkn}$ for any other geodesic simply by multiplying by $e^{-ik\Phi^\theta(0)}$.

In Hughes et al.~\cite{Hughes:2021exa}, they computed amplitudes $Z_{\lmkn}$ for a family of fiducial geodesics. To perform adiabatic evolutions, they then took inspiration from the geodesic form~\eqref{geodesic} (or the analogous form for the asymptotic wave) by replacing $Z_{\lmkn}$ with $\hat Z_{\lmkn} = Z_{\lmkn}e^{-i\Phi_{mkn}(0)}$ in our Eq. (2) and then defining adiabatically evolving phases $\Phi^A(t)$ as the solution to our Eq.~(1) with vanishing initial values; see their (4.1)--(4.2) and (3.18). By treating $\Phi^\theta(0)$ (or an equivalent quantity) as a function of $I_A$ and evolving it along with $I_A$, they then effectively included some post-adiabatic information in their waveform.

Such a scheme is equivalent to ours at adiabatic order, but it does not fit neatly into the two-timescale scheme. In the two-timescale scheme, rather than solving field equations with geodesic sources and then feeding their outputs into an adiabatic evolution, one works directly with the expansion~\eqref{generic} throughout the calculation; the underlying assumption is that the physical, evolving system is periodic with respect to each of the physical, evolving phases (rather than being triperiodic with respect to $t$ as is the case for a geodesic source). Given the expansion~\eqref{generic}, one then solves field equations with the source $T^{\mu\nu}_{mkn}$ rather than the source $\hat T^{\mu\nu}_{mkn}$. Since $\hat T^{\mu\nu}_{mkn} = T^{\mu\nu}_{mkn}$ for a fiducial geodesic, {\it the solutions to these field equations are precisely the $Z_{\lmkn}$ one would obtain for the fiducial geodesic}. But one need not invoke geodesics of any kind, fiducial or otherwise, when solving the field equations, and one entirely loses the association with geodesics at post-adiabatic orders. Instead one works with the genuine, evolving parameters and phases throughout, and the evolving inspiral and waveform are determined entirely by the evolution equations for $I_A$ and $\Phi^A$.

However, suppose one has already computed $Z_{\lmkn}$ for a given family of geodesics. One can always use this family as input for the two-timescale expansion by appropriately choosing the origin of the $\Phi^A$ phase space. For example, in the $5$PN-$e^{10}$ calculations we use geodesics with $r_{\rm p}(0)=r_{\rm max}$, $\theta_{\rm p}(0)=\pi/2$, and $\frac{d\theta_{\rm p}}{dt}(0) < 0$. We can then define $\Phi^A$ such that $\Phi^A=0$ at $r_{\rm p}=r_{\rm max}$, $\theta_{\rm p}=\pi/2$, and $\varphi_{\rm p}=0$. The Teukolsky amplitudes $Z_{\lmkn}$ obtained for the geodesics with $r_{\rm p}(0)=r_{\rm max}$ and $\theta_{\rm p}(0)=\pi/2$ then precisely correspond to the amplitudes in the two-timescale expansion. Note that this choice of origin in phase space does not restrict the initial conditions for Eq.~(1), which are freely specified.

\section{Passage through resonance}
Next, we briefly describe the derivation of the evolution equations 
of orbital parameters to compute the resonant jumps. 
This requires the oscillatory pieces of the first-order GSF in terms of the 
phase $\Phi^A$. 
Below we focus our attention exclusively on the dissipative sector of the evolution equations, as we do in the body of the letter. 
The conservative sector is discussed in Ref.~\cite{Fujita:2016igj}.

The evolution equations are most easily extracted from $dJ^{\rm osc}_A/dt$, where the alternative orbital parameters $J^{\rm osc}_A=\{{\hat E},\, {\hat L},\,{\hat Q}\}$ are the (specific) orbital energy, azimuthal angular momentum, and Carter constant, defined as ${\hat E} = -u_t$, ${\hat L} = u_\phi$, 
and $ {\hat Q} = u_\alpha u_\beta K^{\alpha\beta} [= {\hat C} + (a {\hat E} - {\hat L})^2]$, where $K^{\alpha\beta}$ is the Killing tensor of Kerr spacetime 
(see, e.g., Sec. 2 in Ref.~\cite{Sago:2005fn}). 
The rates of change of these quantities are (see Appendix. C of Ref.~\cite{Isoyama:2018sib})
\begin{equation}\label{dEdtau}
\frac{dJ^{\rm osc}_A}{dt} 
= -\frac{1}{u^t}
\left(
 \frac{\partial J^{\rm osc}_A}{\partial u_{\alpha}}
 \right)_{x}
 \left(
 \frac{\partial H^{(1)} }{\partial x^{\alpha}}
 \right)_{\!J} \equiv F_A \,,
\end{equation}
using $u_{\alpha} = u_{\alpha}(x,\,J^{\rm osc})$~\cite{Schmidt:2002qk} 
and $d/dt = (d/d\tau)/u^t$ with the proper time $\tau$ 
compatible with the background Kerr metric. 
$H^{(1)}$ is the dissipative (i.e., time-antisymmetric) piece of the particle's perturbed Hamiltonian 
to describe the self-acceleration, defined by  
\begin{equation}\label{H_rad}
H^{(1)}
\equiv 
- \frac{1}{2} {h}^{\alpha \beta}_{\text{rad}} \, u_{\alpha} u_{\beta} \,, 
\end{equation}
where  
$h^{\rm rad}_{\alpha\beta}
=
\frac{1}{2}\left(h^{\rm ret}_{\alpha\beta} - h^{\rm adv}_{\alpha\beta}\right)$ 
is the (half-retarded-minus-half-advanced) radiative piece of the metric perturbation, 
and all quantities are evaluated on the orbit. 

Similarly to Eq.~\eqref{generic}, we can expand $h^{\rm rad}_{\alpha\beta}$ 
in the Fourier modes
\begin{equation}\label{hrad}
h^{\rm rad}_{\alpha\beta} 
= 
\sum_{mkn}h^{{\rm rad},mkn}_{\alpha\beta}(J^{\rm res}_A,r,\theta)\,
e^{-i\Phi_{mkn}+im\varphi}, 
\end{equation}
where we have evaluated $J^{\rm osc}_A$ at $J^{\rm res}_A\equiv J^{\rm osc}_A(t_{\rm res})$ inside the metric perturbation, using the fact that $J^{\rm osc}_A$ only changes by an amount $\sim \epsilon^{1/2}$ across the resonance, 
and hence it effectively remains constant. 
On the orbit we have $\varphi=\varphi_{\rm p} = \Phi^\varphi + \Delta \varphi(\Phi^r,\Phi^\theta)$, such that $\Phi^\varphi$ cancels out in Eq.~\eqref{hrad} and in Eq.~\eqref{H_rad}. $r_{\rm p}$, $\theta_{\rm p}$, and $u^\alpha$ are likewise independent of $\Phi^\varphi$ and biperiodic in $(\Phi^r,\Phi^\theta)$, 
such that 
\begin{equation}\label{dJdt}
\frac{dJ^{\rm osc}_A}{dt} = \sum_{mkn,k'n'}F^{mkn,k'n'}_A(J^{\rm res}_B)e^{i(n'-n)\Phi^r+i(k'-k)\Phi^\theta}.
\end{equation}
Here, $mkn$ are the mode numbers from Eq.~\eqref{hrad}, and $k'n'$ 
are those from the Fourier expansion of the $(r_{\rm p},\theta_{\rm p}, u^\alpha)$ dependence. 

Since $F_A$ is biperiodic in $(\Phi^r,\Phi^\theta)$, we can write it as $F_A= \sum_{{\sf k}_r,{\sf k}_\theta}F^{{\sf k}_r,{\sf k}_\theta}_A(J^{\rm res}_B)e^{i\Phi_{{\sf k}_r,{\sf k}_\theta}}$, with $\Phi_{{\sf k}_r,{\sf k}_\theta} 
\equiv 
( {\mathsf k}_{r}\, \Phi^{r} + {\mathsf k}_{\theta}\,\Phi^{\theta} )$ and with
coefficients $F^{{\sf k}_r,{\sf k}_\theta}_A\equiv \frac{1}{(2\pi)^2}\oint d^2\Phi\,F_A e^{-i\Phi_{{\sf k}_r,{\sf k}_\theta}}=\sum_{mkn}\sum_{\substack{k'=k+{\sf k}_\theta\\ n' = n+{\sf k}_r}} F^{mkn, k'n'}_A(J^{\rm res}_B)$. The modes $({\sf k}_r,{\sf k}_\theta)=(0,0)$, 
whether the orbit is away from resonance or on resonance, do not oscillate. 
At resonance, all modes $({\sf k}_r,{\sf k}_\theta)=(s\beta^r,-s\beta^\theta)$ satisfying $\Omega_r \beta^r - \Omega_\theta \beta^\theta = 0$ also
become stationary. We then have 
\begin{equation}\label{GSF-res}
F^{{\rm res},s}_A 
\equiv \sum_{mkn}
F^{mkn,k - s\beta^\theta, n + s\beta^r}_A(J^{\rm res}_B).
\end{equation}
Writing $I^{\rm osc}_A = I^{\rm osc}_A(J^{\rm osc}_B)$, 
where the one-to-one relationship is the geodesic one 
(see, e.g., Appendix~B in Ref.~\cite{Schmidt:2002qk}), we obtain 
$G^{{\rm res},s}_A = \frac{\partial I_A}{\partial J_B}F^{{\rm res},s}_B$.

The coefficients in Eq.~\eqref{GSF-res} can be expressed in terms of the Teukolsky mode amplitudes $Z_{\ell mkn}$, as in Eq. (8) in the body of our letter, by expressing the expansion~\eqref{hrad} 
in terms of a radiative Green's function $G^{\rm rad}_{\alpha\beta\alpha'\beta'}(x,x')$, 
as in Appendix~A of Sago et al.~\cite{Sago:2005fn}.  
In broad strokes, we have 
$h^{\rm rad}_{\alpha\beta}(x) = 
\int G^{\rm rad}_{\alpha\beta\alpha'\beta'}(x,x')T^{\alpha'\beta'}(x')dV'$ and 
$G^{\rm rad}_{\alpha\beta\alpha'\beta'} \sim 
\sum_{\ell mkn}\left[\Pi^+_{\ell mkn}(x)\bar\Pi^+_{\ell mkn}(x') 
+ \Pi^-_{\ell mkn}(x)\bar\Pi^-_{\ell mkn}(x')\right]$, 
where $\Pi^\pm_{\ell mkn} = (\Pi^\pm_{\ell mkn})_{\alpha \beta}$ are constructed 
from homogeneous solutions to the Teukolsky equation, 
with $\Pi^{+/-}_{\ell mkn}$ regular at infinity / the horizon
(and an overbar denoting the complex conjugate of a quantity). 
The integral of $\bar\Pi^\pm_{\ell mkn}(x')T^{\alpha'\beta'}(x') 
\sim (\bar\Pi^\pm_{\ell mkn})_{\alpha' \beta'} u^{\alpha'} u^{\beta'}$ 
can be expressed in terms of $Z^\pm_{\ell mkn}$, leaving terms of the form 
$h^{\rm rad}_{\alpha\beta}(x)\sim Z^\pm_{\ell mkn}\Pi^\pm_{\ell mkn}$. 
When we evaluate $H^{(1)}(x) \sim h^{\alpha \beta}_{(\text{rad})}(x) \, u_{\alpha} u_{\beta} 
\sim (\Pi^\pm_{\ell mkn})^{\alpha \beta} u_{\alpha} u_{\beta}$ on the worldline 
and integrate against $e^{-i\Phi_{k'n'}}$ to construct $F^{mkn,k'n'}_A$, 
the integral can again be expressed in terms of ${\bar Z}^{\pm}_{\ell mkn}$ 
(for a given ${\sf k}_r$ and ${\sf k}_\theta$), leading to Eq. (8).  
Very similar calculations in the case of ${\sf k}_r = 0 = {\sf k}_\theta$ 
are detailed in, e.g., Sago et al.~\cite{Sago:2005gd}; 
see the steps in going from Eq. (4.3) to (4.6) in that reference, 
and Sec. 8 and 9 of Drasco et al.~\cite{Drasco:2005is} (in the scalar toy-model case).

Equation (8) can also be obtained from calculations 
of `resonantly enhanced (or diminished) fluxes' 
in geodesic snapshots~\cite{Flanagan:2012kg,Isoyama:2013yor,Isoyama:2018sib}.
Unlike our method, those calculations used the snapshot phase 
$\Phi^A(t) = \Omega^A t + \Phi^A(0)$ and explicitly took a long-time average 
of Eq.~\eqref{dJdt} 
(such that $\Phi^A(0)$ created an `offset'-phase dependence 
in the time-averaged fluxes at resonances~\cite{Grossman:2011im}).
But prior to taking the long-time average, we can immediately promote the snapshot phase $\Phi^A(t)$ 
in the existing calculations  
to the adiabatic phase of our two-timescale scheme, 
leaving their intermediate results unaffected. 
Indeed, the essence of Eq. (8) in the letter can be straightforwardly 
extracted from, for example, Eq. (54) 
in Isoyama et al.~\cite{Isoyama:2018sib};  
see also Eq. (4.6) in Sago et al.~\cite{Sago:2005gd} and 
Eqs. (5.1) and (5.2) in Ruangsri \& Hughes~\cite{Ruangsri:2013hra}.

\begin{figure}[tbh]
    \includegraphics[angle=0,width=\columnwidth]{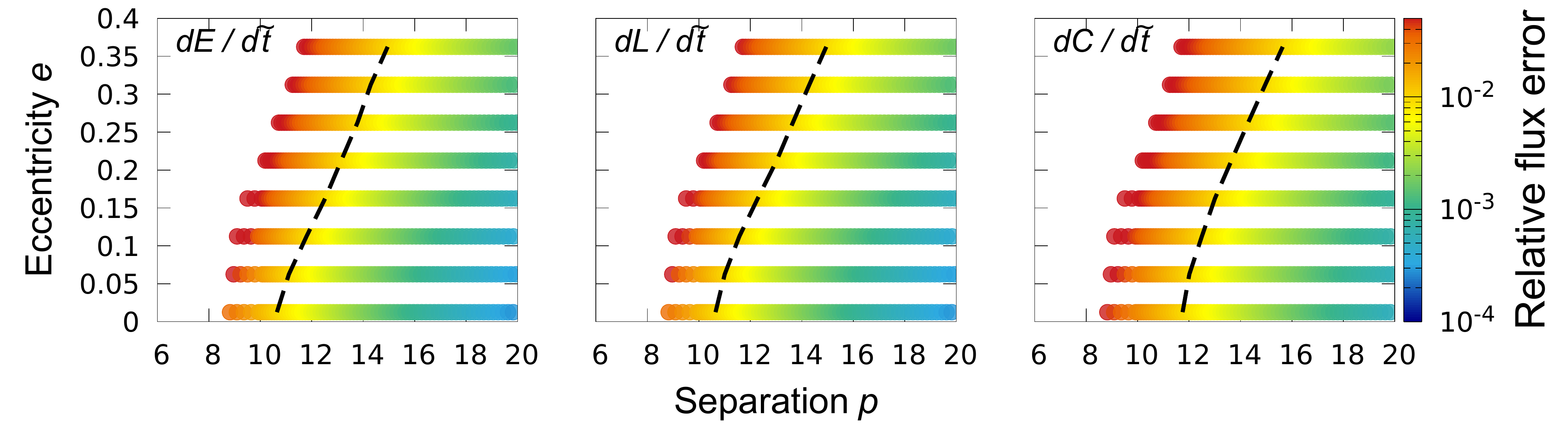}
	\caption{Error of $5$PN-$e^{10}$ fluxes relative to numerical flux data 
	in the worst case of $q = -0.9$, $\iota \approx 20^{\circ}$. 
	This is the same parameters used in the right panel of FIG. 2. 
	We show the errors of $dE / d{\tilde t}$ ({\textit {Left}}), 
	$dL / d{\tilde t}$ ({\textit {Middle}}), 
	and $dC / {d {\tilde t}}$ ({\textit {Right}}), respectively. 
	The dashed curves indicate an error of $\approx 1.0 \times 10^{-2}$.}
	\label{fig:rel_err_q-0.90inc20}
\end{figure}

\begin{figure}[tbh]
    \includegraphics[angle=0,width=\columnwidth]{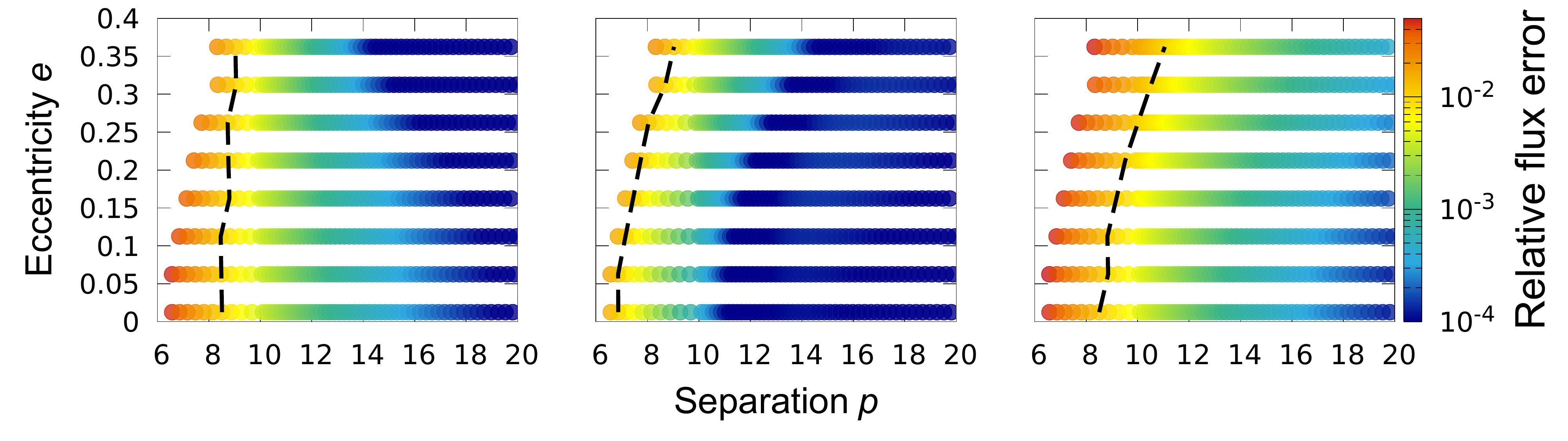}
	\caption{The same figures as FIG.~\ref{fig:rel_err_q-0.90inc20}
	[the errors of $dE / d{\tilde t}$ ({\textit {Left}}), $dL / d{\tilde t}$ ({\textit {Middle}}),  
	and $dC / d{\tilde t}$ ({\textit {Right}})], 
	but in the best case of $q = +0.9$, $\iota \approx 80^{\circ}$.
    This is the same parameters used in the left panel of FIG. 2.}
	\label{fig:rel_err_q0.90inc80}
\end{figure}

{\section{Relative flux errors of other orbital parameters}}
In FIG. 2 of this letter, we showed a sample of the relative flux error of $p$.  
Here, for completeness, we plot additional flux errors of 
`traditional' orbital parameters 
$\{{E},\, {L},\,{C}\} (\equiv \{\mu {\hat E},\, \mu {\hat L},\, \mu^2 {\hat C}\})$ 
in FIG.~\ref{fig:rel_err_q-0.90inc20} for the worst case $q = -0.9$, $\iota \approx 20^{\circ}$, 
and in FIG.~\ref{fig:rel_err_q0.90inc80}
for the best case $q = +0.9$, $\iota \approx 80^{\circ}$;  
recall that the mapping between $\{{E},\, {L},\,{C}\}$ and $\{p,\,e,\,\iota\}$ is one-to-one. 
These errors are qualitatively similar to those in $dp / d{\tilde t}$.

\section{Dephasing between adiabatic and first post-adiabatic models}

As an example of the dephasing time between the $5$PN-$e^{10}$ 
and numerical adiabatic models, here we report a specific comparison for an equatorial EMRI (the only type for which there are many numerical adiabatic evolutions). We choose an example for which data is available in Ref.~\cite{Fujita:2020zxe} and which has parameters close to those of our sample generic, inclined EMRI in the letter,
with masses and Kerr spin $(\mu,\,M,\,q) = (10 M_{\odot},\,10^6M_{\odot},\,0.9)$, 
initial phases $\Phi^A(0) = 0$, and initial orbital parameters $(p(0),\,e(0)) = (10.0,\,0.2)$. 
In this case we find a phase difference 
$(\delta \Phi_r,\,\delta \Phi_{\varphi}) 
\approx (0.60,\,0.77)$ after the first $\approx 8$ weeks (${\tilde t} = 10.0M$) of evolution. 

Next we justify our estimate that this time is comparable to the time over which an 
adiabatic waveform will maintain phase coherence with a post-adiabatic one.
At first post-adiabatic order, away from resonances, the evolution equations can be put in the form
\begin{equation}
\frac{d\Phi^A}{d\tilde t} = \eta^{-1}\Omega^A(I_B) \quad\text{and}\quad \frac{dI_A}{d\tilde t} = G^{(0)}_A(I_B) + \eta G^{(1)}_A(I_B).
\end{equation}
These equations admit an asymptotic solution of the form
\begin{align}
\Phi^A &= \eta^{-1}\Phi^A_{(0)}(\tilde t) + \eta^0 \Phi^A_{(1)}(\tilde t) + O(\eta),\\
\Omega^A &= \Omega^A_{(0)}(\tilde t) + \eta \Omega^A_{(1)}(\tilde t) + O(\eta^2),
\end{align}
where $\frac{d\Phi^A_{(n)}}{d\tilde t} = \Omega^A_{(n)}(\tilde t)$.

An adiabatic evolution captures $\Phi^A_{(0)}$ but omits $\Phi^A_{(1)}$. This implies that for each mode of the waveform, it omits a post-adiabatic phase correction $\Phi^{(1)}_{mkn} = \eta \int_0^t \omega^{(1)}_{mkn}(\eta t')dt'$, which is bounded by
\begin{equation}\label{error}
|\Phi^{(1)}_{mkn}| < \eta \omega^{(1){\rm max}}_{mkn} t,
\end{equation}
where $\omega^{(1){\rm max}}_{mkn} \equiv \max_{0<t'<t}{|\omega^{(1)}_{mkn}(\eta t')|}$. 

Equation~\eqref{error} is an upper bound on the error. We can limit it to an error tolerance $N$ by restricting to a time interval
\begin{equation}
t < N/(\eta \omega^{(1){\rm max}}_{mkn}).
\end{equation}
A typical case might be $N \sim 0.1$~rad, a mass ratio $\eta = 10^{-5}$, and $\omega^{(1){\rm max}}_{mkn}\sim 10^{-2}$~rad/s (i.e., frequencies in the LISA band, and the typical post-adiabatic correction to the frequency of the innermost stable circular orbit, caused by the first-order conservative GSF~\cite{Barack:2009ey,LeTiec:2011dp,Isoyama:2014mja,vandeMeent:2016hel,Lynch:2021ogr}). This implies a time interval $t\lesssim 16$ weeks.


